\PassOptionsToPackage{table,xcdraw}{xcolor}
\documentclass[sigconf, authorversion]{acmart}

\usepackage[a-1b]{pdfx}
\usepackage{hyperref}
\usepackage{graphicx}
\usepackage{subcaption}
\usepackage{balance}  
\usepackage{multicol}
\usepackage{vcell}
\usepackage{tabularx,booktabs}
\newcolumntype{a}{>{\hsize=.08\hsize}X}
\newcolumntype{b}{>{\hsize=.13\hsize}X}
\newcolumntype{d}{>{\hsize=.35\hsize}X}
\newcolumntype{s}{>{\hsize=.45\hsize}X}
\newcolumntype{t}{>{\hsize=.45\hsize}X}
\newcolumntype{u}{>{\hsize=.5\hsize}X}
\newcolumntype{v}{>{\hsize=.01\hsize}X}
\usepackage{url}

\usepackage{breakurl}

\usepackage{tikz}
\usetikzlibrary{shapes}
\tikzstyle{block} = [rectangle, draw,
    text width=5em, text centered, rounded corners, minimum height=2em]

\tikzstyle{block2} = [rectangle, draw,
    text width=6em, text centered, rounded corners, minimum height=2em]

\tikzstyle{bigblock} = [rectangle, draw,
    text width=10em, text centered, rounded corners, minimum height=10em]

\usepackage{tikz}

\usepackage{pgfplots}
\usepackage{pgfplotstable}
\usepgfplotslibrary{
        groupplots,
        statistics
    }

\pgfplotsset{select coords between index/.style 2 args={
    x filter/.code={
        \ifnum\coordindex<#1\fi
        \ifnum\coordindex>#2\fi
    }
}}

\usepgfplotslibrary{external}
\usepackage{pgfplotstable}
\pgfplotstableset{
col sep=semicolon
}

\usetikzlibrary{external}
\tikzsetexternalprefix{tikz/}
\tikzset{external/force remake} 

\AtBeginDocument{%
  \providecommand\BibTeX{{%
    \normalfont B\kern-0.5em{\scshape i\kern-0.25em b}\kern-0.8em\TeX}}}

\copyrightyear{2021} 
\acmYear{2021} 
\setcopyright{acmlicensed}\acmConference[ICPE '21]{Proceedings of the 2021 ACM/SPEC International Conference on Performance Engineering}{April 19--23, 2021}{Virtual Event, France}
\acmBooktitle{Proceedings of the 2021 ACM/SPEC International Conference on Performance Engineering (ICPE '21), April 19--23, 2021, Virtual Event, France}
\acmPrice{15.00}
\acmDOI{10.1145/3427921.3450242}
\acmISBN{978-1-4503-8194-9/21/04}

\begin{document}
 \fancyhead{}
\title{ESPBench: The Enterprise Stream Processing Benchmark}

\author{Guenter Hesse}

\orcid{0000-0002-7634-3021}
\affiliation{%
  \institution{Hasso Plattner Institute\\University of Potsdam}
  \streetaddress{August-Bebel-Str. 88}
  \city{Potsdam}
  \state{Germany}
  \postcode{14482}
}
\email{guenter.hesse@hpi.de}

\author{Christoph Matthies}
\affiliation{%
  \institution{Hasso Plattner Institute\\University of Potsdam}
  \streetaddress{August-Bebel-Str. 88}
  \city{Potsdam}
  \state{Germany}
  \postcode{14482}
}
\email{christoph.matthies@hpi.de}

\author{Michael Perscheid}
\affiliation{%
  \institution{Hasso Plattner Institute\\University of Potsdam}
  \streetaddress{August-Bebel-Str. 88}
  \city{Potsdam}
  \state{Germany}
  \postcode{14482}
}
\email{michael.perscheid@hpi.de}

\author{Matthias Uflacker}
\affiliation{%
  \institution{Hasso Plattner Institute\\University of Potsdam}
  \streetaddress{August-Bebel-Str. 88}
  \city{Potsdam}
  \state{Germany}
  \postcode{14482}
}
\email{matthias.uflacker@hpi.de}

\author{Hasso Plattner}
\affiliation{%
  \institution{Hasso Plattner Institute\\University of Potsdam}
  \streetaddress{August-Bebel-Str. 88}
  \city{Potsdam}
  \state{Germany}
  \postcode{14482}
}
\email{hasso.plattner@hpi.de}
% \renewcommand{\shortauthors}{B. Trovato et al.}
\renewcommand{\shortauthors}{G. Hesse et al.}
\begin{abstract}
Growing data volumes and velocities in fields such as Industry 4.0 or the Internet of Things have led to the increased popularity of data stream processing systems.
Enterprises can leverage these developments by enriching their core business data and analyses with up-to-date streaming data.
Comparing streaming architectures for these complex use cases is challenging, as existing benchmarks do not cover them.
\emph{ESPBench} is a new enterprise stream processing benchmark that fills this gap.
We present its architecture, the benchmarking process, and the query workload.
We employ ESPBench on three state-of-the-art stream processing systems, \emph{Apache Spark}, \emph{Apache Flink}, and  \emph{Hazelcast Jet}, using provided query implementations developed with \textit{Apache Beam}.
Our results highlight the need for the provided ESPBench toolkit that supports benchmark execution, as it enables query result validation and objective latency measures.
\end{abstract}





\maketitle

\section{Introduction}
\label{introduction}

\begin{sloppypar}

The need to process growing data volumes has led to the increased importance of data stream processing approaches.
While in 2016, the music streaming service \emph{Spotify} handled 1.5M\,events/second, this number had dramatically increased to 8M\,events/second in 2018~\cite{twitterspotify}.
Even businesses less involved with the digital world face high amounts of data, e.g., those from the manufacturing domain.
For example, an ultrasonic sensor production plant creates about 170\,GB of data per day~\cite{nagy2018role}.
Manufacturing equipment such as a single saw can generate 50,000 messages or 1.2G\,B of data on a daily basis~\cite{crateio}.
Injection molding machines even produce up to multiple terabytes of sensor data in 24 hours~\cite{DBLP:conf/gi/HuberVN16}.
These developments highlight the large data volumes companies are facing today as well as the high rate of growth.

Various new data stream processing systems (DSPSs), which are leveraged for analyzing continuously generated data, have been developed recently~\cite{DBLP:conf/icpads/HesseL15}.
This increased choice has led to more options for DSPS users, which raises the question of how to identify the DSPS that best satisfies current and future demands.
Performance benchmarks offer solutions for this challenge as they reveal performance differences between systems or configurations.
Currently, there is no satisfying benchmark for DSPSs that comprises:
\end{sloppypar}

\begin{itemize}
  \item Integration of existing, traditional business data, such as production orders or customer information~\cite{DBLP:conf/data/HesseSMU19}
  \item Satisfying tool support (data ingestion, query result validation, objective performance result calculation, automation)
  \item Coverage of the core DSPS functionalities, e.g., windowing and transformation capabilities.
\end{itemize}

\begin{table*}[!htb]
\caption{Jim Gray's~\cite{DBLP:books/mk/Gray93} design criteria for domain-specific database benchmarks as applied to data stream processing systems}
\vspace{-0.5em}
\label{tab:designcriteria}
\begin{tabular}{lll}
\toprule
Criteria                     & Description                                                                                                                                                                                                                                         & Applied to DSPS benchmarks                                                                                                                                                                                                                                                                                                                                                                      \\
\hline
\vcell{\textbf{Relevance}}   & \vcell{\begin{tabular}[b]{@{}l@{}}\textcolor[rgb]{0.2,0.2,0.2}{Typical operations of the problem domain}\\\textcolor[rgb]{0.2,0.2,0.2}{need to be tested}\end{tabular}}                 & \vcell{\begin{tabular}[b]{@{}l@{}}\begin{tabular}{@{\labelitemi\hspace{\dimexpr\labelsep+0.5\tabcolsep}}l}\textcolor[rgb]{0.2,0.2,0.2}{Queries cover core functionalities of DSPSs}\\\textcolor[rgb]{0.2,0.2,0.2}{Queries validated with industry}\\\textcolor[rgb]{0.2,0.2,0.2}{Design represents real-world settings}\\\textcolor[rgb]{0.2,0.2,0.2}{Data rates configurable}\end{tabular}\\[-1em]\textcolor[rgb]{0.2,0.2,0.2}{}\end{tabular}}  \\[-\rowheight]
\printcelltop                & \printcelltop                                                                                                                                                                                                                                       & \printcelltop   \\
\vcell{\textbf{Portability}} & \vcell{\begin{tabular}[b]{@{}l@{}}\textcolor[rgb]{0.2,0.2,0.2}{Easy to use on different systems or architectures~}\textcolor[rgb]{0.2,0.2,0.2}{}\\\textcolor[rgb]{0.2,0.2,0.2}{}\end{tabular}} & \vcell{\begin{tabular}[b]{@{}l@{}}\begin{tabular}{@{\labelitemi\hspace{\dimexpr\labelsep+0.5\tabcolsep}}l}\textcolor[rgb]{0.2,0.2,0.2}{No restrictions on SUT architecture specifics by toolkit}\\\textcolor[rgb]{0.2,0.2,0.2}{System-independent query definitions}\\[1.7em]\end{tabular}\end{tabular}}                                                                                                                        \\[-\rowheight]
\printcelltop                & \printcelltop                                                                                                                                                                                                                                       & \printcelltop                                                                                                                                                                                                                                                                                                                                                                                            \\
\vcell{\textbf{Scalability}} & \vcell{\begin{tabular}[b]{@{}l@{}}\textcolor[rgb]{0.2,0.2,0.2}{Applicable to both, single node systems and~}\\\textcolor[rgb]{0.2,0.2,0.2}{}\textcolor[rgb]{0.2,0.2,0.2}{scale-out systems with multiple nodes;}\\\textcolor[rgb]{0.2,0.2,0.2}{}\textcolor[rgb]{0.2,0.2,0.2}{benchmark should be scalable to larger systems}\end{tabular}}                        & \vcell{\begin{tabular}[b]{@{}l@{}}\begin{tabular}{@{\labelitemi\hspace{\dimexpr\labelsep+0.5\tabcolsep}}l}\textcolor[rgb]{0.2,0.2,0.2}{Data rates configurable/scalable}\\\textcolor[rgb]{0.2,0.2,0.2}{Scale-out scenarios supported by benchmark tools}\\[1em]\end{tabular}\end{tabular}}                                                                                                                                         \\[-\rowheight]
\printcelltop                & \printcelltop                                                                                                                                                                                                                                       & \printcelltop                                                                                                                                                                                                                                                                                                                                                                                            \\
\vcell{\textbf{Simplicity}}  & \vcell{\begin{tabular}[b]{@{}l@{}}\textcolor[rgb]{0.2,0.2,0.2}{Easy to understand and easy to use/implement~}\\\textcolor[rgb]{0.2,0.2,0.2}{}\textcolor[rgb]{0.2,0.2,0.2}{to ensure result creditbility}\end{tabular}}                                                                                                                                                                  & \vcell{\begin{tabular}[b]{@{}l@{}}\begin{tabular}{@{\labelitemi\hspace{\dimexpr\labelsep+0.5\tabcolsep}}l}\textcolor[rgb]{0.2,0.2,0.2}{Automation of entire benchmark process}\\\textcolor[rgb]{0.2,0.2,0.2}{Publication of usable example query implementation}\\\textcolor[rgb]{0.2,0.2,0.2}{Tool support for essential benchmark functions, e.g.,}\\\end{tabular}\\\begin{tabular}[b]{@{}l@{}}\textcolor[rgb]{0.2,0.2,0.2}{query result validation and data stream generation}\end{tabular}\end{tabular}}                                                                                                              \\[-\rowheight]
\printcelltop                & \printcelltop                                                                                                                                                                                                                                       & \printcelltop                                                                                                                                                                                                                                                                                                                                                                                            \\
\bottomrule
\end{tabular}
\vspace{-1em}
\end{table*}

We close this gap with a benchmark for enterprise stream processing architectures.
The contributions are as follows:
\begin{itemize}
  \item
  We propose \emph{ESPBench}, an enterprise stream processing benchmark.
  It includes a toolkit for data ingestion, query result validation, benchmark result calculation, and automation.
  \item
  We present an example implementation of the ESPBench queries using Apache Beam~\cite{beam}, an abstraction layer for defining data processing applications, which can be executed on any of the currently more than ten supported DSPSs~\cite{beamcapa}.
  \item
  We conduct an experimental evaluation, benchmarking three state-of-the-art stream processing systems with the Apache Beam example implementation with the objective to validate the concepts and tools of ESPBench.
\end{itemize}

To allow result reproduction and the use of ESPBench, we published all artifacts~\cite{guenter_hesse_2020_4010167} and created a public repository\footnote{\url{https://github.com/guenter-hesse/ESPBench}}.
The remainder of this paper is structured as follows:
Section~\ref{sec:ESPBench} presents ESPBench.
We describe the benchmark scenario and its architecture, the benchmark process, the input data, and the queries.
Section~\ref{sec:hbvali} introduces the validation setup for ESPBench, illustrating the systems under test and the benchmarking landscape.
Subsequently, we discuss the benchmark results in the experimental evaluation section, followed by an outline of threats to validity.
Sections~\ref{sec:lessonslearned} and~\ref{sec:relatedwork} elaborate on the lessons learned and highlight related work.
The last section concludes and gives an outlook on future work.
\vspace{-0.5em}
\section{The ESPBench Benchmark}
\label{sec:ESPBench}
This section introduces the developed ESPBench, a performance benchmark with comprehensive tool support; covering all core functionalities of DSPSs, including the combination of streaming data with structured business data.
It builds upon the ideas presented in~\cite{DBLP:conf/debs/HesseMRU17} and~\cite{DBLP:conf/tpctc/HesseRMLKU17}.
We start by presenting the overall design objectives associated with ESPBench.
Afterward, we give an overview of the benchmark scenario, its architecture as well as workflow, and present the employed input data and queries.
\vspace{-0.5em}
\subsection{Design Objectives}
\label{subsec:designobjectives}

One of the most influential works on design principles for benchmarking is \emph{The Benchmark Handbook for Database and Transaction Processing Systems} by Jim Gray~\cite{DBLP:books/mk/Gray93}.
He defines four criteria a domain-specific benchmark has to meet to be considered useful.
Although the work is from the early 90s, it shows major overlaps to newer publications of its kind, such as the work by Karl Huppler~\cite{DBLP:conf/tpctc/Huppler09} or v. Kistowski et al.~\cite{DBLP:conf/wosp/KistowskiAHLHC15}.
The four criteria defined by Gray are described in Table~\ref{tab:designcriteria}.
We translated these rather general aspects to the requirements of DSPS benchmarks.
These applied design principles build the foundation for ESPBench, i.e., they are taken into account for all design decisions and satisfied by ESPBench.
\vspace{-0.5em}
\subsection{Benchmark Scenario}
\label{subsec:scenario}
The benchmark scenario sketches the management of a production plant.
It is inspired by the 2012 \emph{Grand Challenge} published at the \textit{Distributed and Event-Based Systems (DEBS)} conference~\cite{DBLP:conf/debs/JerzakHFGHS12}.
This challenge is about a company that uses high-tech manufacturing machines equipped with sensors.
The company aims to improve the monitoring and analytical capabilities of its manufacturing processes by  making use of the continuously captured sensor data, e.g., by leveraging stream processing technology.
Processing this data enables for, e.g., faster reactions to unintended system states and the combination of sensor and structured business data, i.e., \emph{vertical integration} in the context of Industry 4.0.
The integration of these traditionally separate data sources reveals additional product insights and allows for  a holistic view of the production process~\cite{DBLP:conf/data/HesseSMU19}.

The employed sensors monitor multiple machine variables, e.g., the power consumption or the state of an additive release valve.
Multiple sensors are installed on a single manufacturing machine and their data are collected using an embedded PC.
The benchmark scenario includes two such machines sending sensor values of the same structure.
\vspace{-1em}

\subsection{Benchmark Architecture}
\label{subsec:ESPBencharch}

\begin{figure}[!htb]
\centering
\includegraphics[width=\columnwidth]{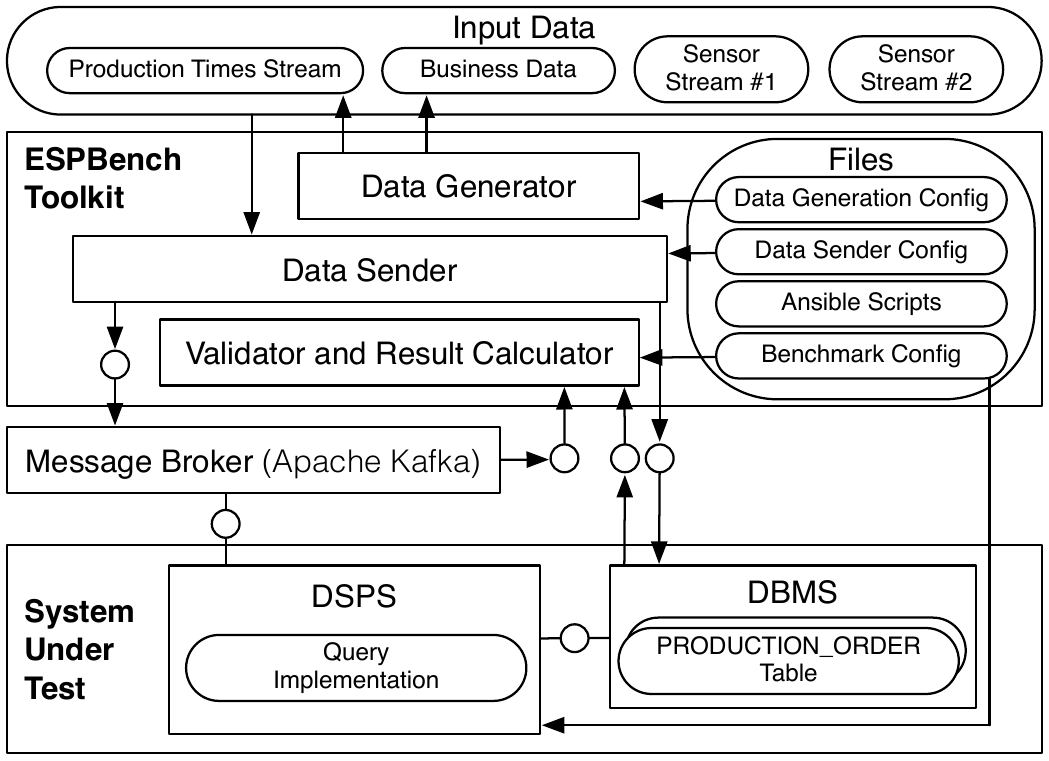}
\caption{Architecture of ESPBench as Fundamental Modeling Concepts (FMC) diagram~\cite{knopfel2005fundamental}}
\label{fig:hbarch}
\vspace{-1em}
\end{figure}

Figure~\ref{fig:hbarch} shows the architecture of ESPBench.
It comprises four major components: \textit{input data, the toolkit, the message broker, and the system under test (SUT)}.
The components labeled as \emph{toolkit} are provided by the benchmark.

The \textit{input data} is stored in \textit{comma-separated values} (CSV) format.
It comprises business data as well as sensor data stored in three different files.
The business data and the production times streaming data are generated by the provided \emph{data generator} as part of the benchmark process.
The other two files can be obtained online, as outlined in the description of ESPBench~\cite{guenter_hesse_2020_4010167}.
Details on the data characteristics and scalability are presented in Section~\ref{subsec:ESPBenchinput}.

The \emph{data sender} is part of the toolkit and responsible for two tasks: the import of the business data into the database management system (DBMS) that is part of the SUT as well as the ingestion of streaming data into the \emph{message broker}.
ESPBench defines a configuration file used to configure several parameters, such as message broker information or the data input rate.

 \begin{figure*}[!htb]
\centering
\includegraphics[width=0.9\textwidth]{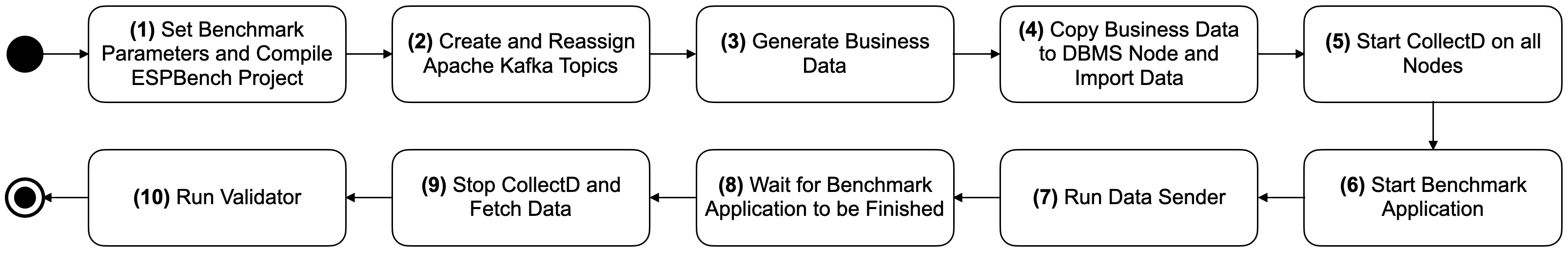}
\caption{ESPBench process visualized as Unified Modeling Language (UML) Activity Diagram~\protect\cite{DBLP:conf/uml/DumasH01}}
\label{fig:activityDiagram}
\vspace{-1em}
\end{figure*}

The \emph{message broker} of ESPBench, \emph{Apache Kafka}~\cite{kreps2011kafka}, is responsible for storing data and represents the interface to the SUT.
ESPBench incorporates a message broker to separate the SUT and the data sender.
This separation allows for realistic and objective performance measurements outside of the SUT, which is important as DSPSs have diverse definitions of latency~\cite{DBLP:conf/icde/KarimovRKSHM18}.
Additionally, changes to the data sender, e.g., launching multiple instances to increase the input rate, does not require adaptions to query implementations.
A potential issue with message brokers identified in~\cite{DBLP:conf/icde/KarimovRKSHM18} is a change to the partitioning during benchmark runs.
We address this point by employing Apache Kafka, which does not perform automated re-partitioning.
The combination of Apache Kafka and a DSPS is comparable to other architectures, both within the domains of performance benchmarking as well as data processing~\cite{DBLP:journals/corr/kafka,lyftarch}.
Thus, this architecture is relevant as it represents real-world environments.

It is crucial to ensure that the message broker does not become a bottleneck since the objective of a benchmark is to analyze the SUT and not any of the tooling components.
Therefore, an input rate that Apache Kafka can manage needs to be configured.
Existing studies on this topic give orientation regarding achievable data input rates for specific settings~\cite{DBLP:journals/corr/kafka}. 
Our studies of multiple manufacturing companies with revenues of at least a billion EUR show that these manageable rates are high enough to represent current environments.

The \emph{SUT} comprises a DSPS and a DBMS in its default setting.
It is responsible for answering the defined benchmark queries.
The benchmark does not impose any scalability restrictions on the SUT, e.g., regarding the cluster size of the DSPS.
\emph{PostgreSQL}~\cite{DBLP:conf/sigmod/StonebrakerR86}, a well-known and widely used DBMS, is the default database of ESPBench.
A change in the DBMS only requires minor adaptions to the toolkit, e.g., to the data sender logic for importing business data into the DBMS.
Alternatively to the SUT default setup, a single system that is able to store the business data and to answer the benchmark queries can be used, reducing communication and data transfer overhead.
While ESPBench provides this flexibility, such a scenario fails to represent most of the current enterprise IT landscapes.

The SUT reads input data from Apache Kafka and writes results back to either Apache Kafka or the DBMS, depending on the query.
The \emph{validator and result calculator} determines the correctness of the query answers and calculates the benchmark results.
The tool determines aggregated results, such as the mean latency and percentiles, as well as single latencies for each output record.
It writes the output to log and CSV files, which can be used for further analyses, like plotting single latencies.
Technically, the tool uses Akka~\cite{akka}, a toolkit for developing distributed applications, also in a streaming fashion.
The validator reads the input data, calculates the query results,  and compares them to the SUT's output.
Furthermore, it computes result latencies, i.e., timestamp differences.
The validator does that by leveraging the Apache Kafka or DBMS timestamps, i.e., the times taken when an input or result record is written to the log of Apache Kafka or the DBMS.
This concept allows for the previously mentioned objective performance measurements outside of DSPSs.
\vspace{-1em}

\subsection{Benchmark Process}
\label{subsec:benchmarkprocess}

The activity diagram in Figure~\ref{fig:activityDiagram} shows the process of ESPBench and sums up the  \emph{Ansible}~\cite{ansible} script, which automates the benchmark steps.
The process steps are exchangeable and can be extended by the user, e.g., to incorporate additional monitoring tools.

In \textit{step (1)}, the benchmark parameters are set and the project is compiled to a fat \emph{Java Archive} (JAR), which contains all dependencies.
The compilation uses \emph{sbt-assembly}~\cite{sbtassembly}, an assembly plugin for the open-source build tool \emph{sbt}~\cite{sbt}.
The setting of parameters is only relevant in scenarios where multiple runs, i.e., multiple rounds of the benchmarking process, are executed and automated.
In this case, the main Ansible benchmarking script is invoked multiple times with the defined parameters for the different runs by an overarching Ansible script.
For instance, distinct runs want to use different Apache Kafka topics to be able to distinguish between runs. 
If there is only a single run, the parameters are read from the configuration files shown in Figure~\ref{fig:hbarch}, which provide default values.

In \textit{step (2)}, the required Apache Kafka topics are created according to the configuration and the naming schema defined by ESPBench.
Topics are the entities in which Apache Kafka organizes and stores data. 
The created topics are then reassigned to assure an even topic distribution across the Apache Kafka brokers.
ESPBench creates topics with one partition, as Apache Kafka only guarantees the correct order of records within a single partition.

In \textit{step (3)}, the Ansible script invokes the \emph{data generator}.
The resulting business data is copied to the DBMS node and imported into the DBMS by the \emph{data sender} tool (\textit{step (4)}).
The Unix daemon \emph{CollectD}~\cite{collectd} is started on all nodes in \textit{step (5)} to gather system data during the benchmark run.
This enables further analyses, such as evaluating differences between configurations regarding memory consumption or CPU utilization.

\textit{Step (6)} starts the benchmark application, i.e., the benchmark query or queries that are to be executed.
After a few seconds wait time for the DSPSs to receive and start the application, the script invokes the \emph{data sender} in \textit{step (7)}, which sends the streaming data to the corresponding Apache Kafka topic(s).
The ESPBench naming convention allows identifying the correct topic names based on the configuration parameters, which eases query implementation.

The \emph{data sender} runs for the configured period of time.
Once it completes, the Ansible script waits until the queries have finished data processing, represented as \textit{step (8)}.
In \textit{step (9)}, CollectD is stopped and the recorded data is transferred to the main node where the Ansible script is executed.
In the last step, \textit{step (10)}, ESPBench invokes the \emph{validator and result calculator}.

\subsection{Input Data}
\label{subsec:ESPBenchinput}

The input data of ESPBench is drawn from two domains: business data and sensor data from manufacturing equipment.
Both types of data are described in the following.

\subsubsection{Sensor Data}
\label{subsubsec:sensordata}

There are two types of data streams that ESPBench incorporates.
The first one is the data set used at the DEBS Grand Challenge 2012, which contains measurements from multiple sensors combined to single records.
There are two machines sending this kind of data.
The record's structure is depicted in Table~\ref{tab:sensdata}.
ESPBench extends the data structure by the column \emph{workplace id}, which is used for combining sensor and business data.
The data includes information from analog as well as binary sensors.

\begin{table}[]
\caption{Data characteristics of the sensor measurement stream (based on~\protect\cite{DBLP:conf/debs/JerzakHFGHS12})}
\label{tab:sensdata}
\small
\begin{tabularx}{\linewidth}{@{}bsu@{}}
\toprule
\# &
Technical Information &
Description
\\
\midrule
1 &
required fixed64 ts &
timestamp
\\
2 &
required fixed64 index &
message index
\\
3 &
required fixed32 mf01 &
electrical power main phase 1
\\
4 &
required fixed32 mf02 &
electrical power main phase 2
\\
5 &
required fixed32 mf03 &
electrical power main phase 3
\\
6-8 &
required fixed32 pc13-pc15 &
anode current drop detection cell 1-3
\\
9-11 &
required uint32 pc25-pc27 &
anode voltage drop detection cell 1-3
\\
12 &
required uint32 res &
unknown
\\
13-18 &
required bool bm05-bm10 &
chemical additive information
\\
19-66 &
optional bool pp01-pp36, pc01-pc06, pc19-pc24 &
unknown
\\
67 & required fixed32 workplaceid & worklace ID
\\
\bottomrule
\end{tabularx}
\vspace{-1em}
\end{table}

Within the sensor data, the column most relevant for ESPBench is \emph{mf01}, the electrical power on main phase one.
It represents the energy consumption of the machines, which is valuable information, e.g., for identifying irregularities in the production process.

\begin{figure}[]
\centering
\includegraphics[width=\columnwidth]{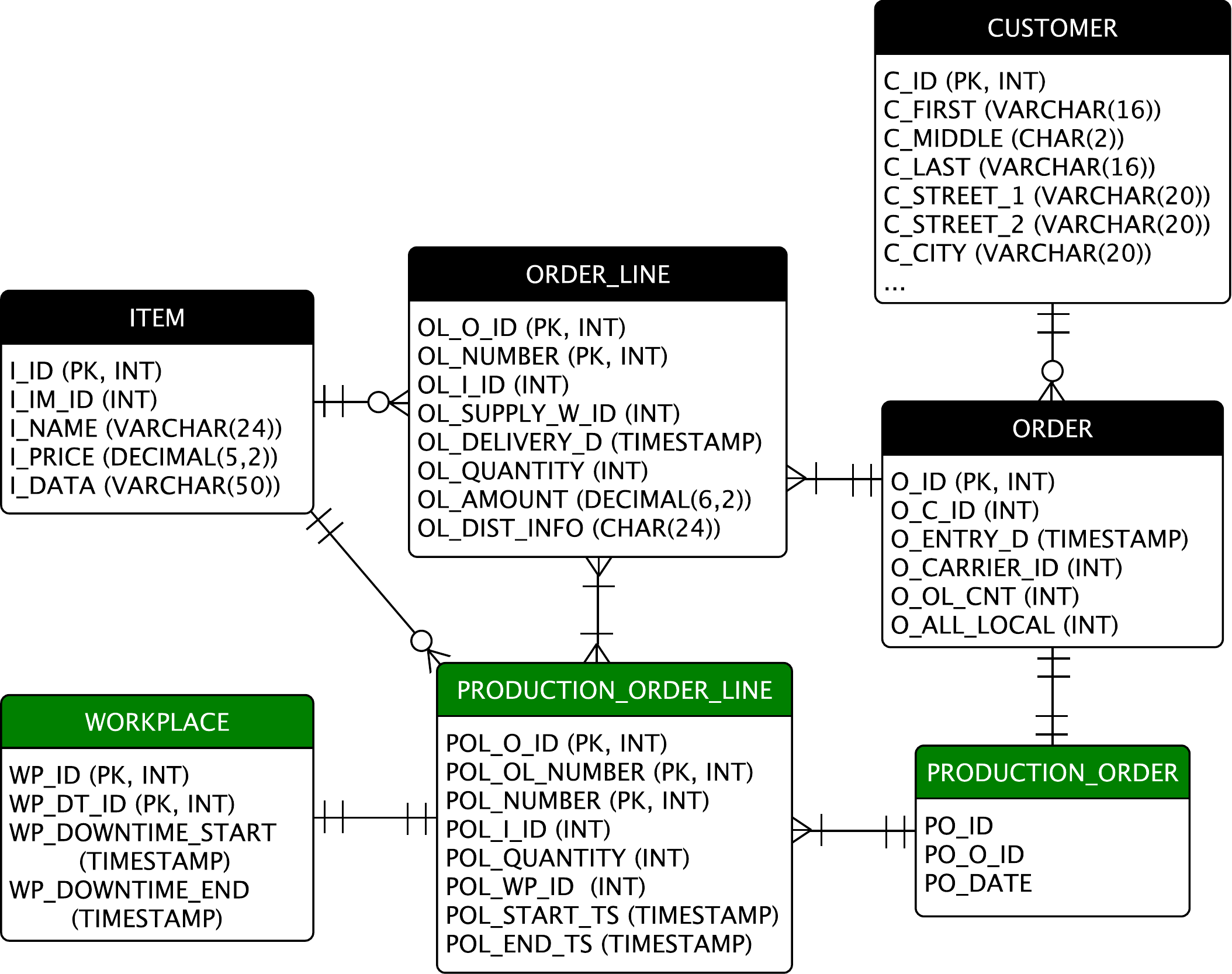}
\caption{ESPBench business data in Crow’s Foot Notation}
\label{fig:serd}
\vspace{-1em}
\end{figure}

The second type of data stream coming from the manufacturing equipment consists of information about the production times that allow combining sensor and business data.
It is not part of the DEBS Grand Challenge, but designed by ESPBench and created by its data generator tool.
The data stream's structure is shown in Table~\ref{tab:stream2_struct}.
It contains the \emph{order id}, \emph{order line number}, \emph{production order line number}, and a column that indicates whether the corresponding product entered or left the workplace.
The structure of the business data that is visualized in Figure~\ref{fig:serd} reveals that the first three columns are the primary key of the \emph{PRODUCTION\_ORDER\_LINE} table and thus, can identify the workplace.
Information about when products entered and left a workplace is needed for time-based vertical data integration, e.g., for linking sensor measurements at manufacturing machines to the product currently being worked on.
Our conducted industry studies showed that this is a broadly adopted practice enterprise settings~\cite{DBLP:conf/dasfaa/HesseMSU19}.

\begin{table}[]
\caption{Data characteristics of the sensor data used for production time determination}
\label{tab:stream2_struct}
\small
\begin{tabularx}{\linewidth}{@{}vlu@{}}
\toprule
\# &
Technical Information &
Description
\\
\midrule
1 &
required uint32 pt\_o\_id &
order id
\\
2 &
required uint32 pt\_ol\_number &
order line number
\\
3 &
required uint32 pt\_pol\_number &
production order line number
\\
4 &
required bool pt\_is\_end &
indicates entering / leaving of workplace
\\
\bottomrule
\end{tabularx}
\vspace{-1em}
\end{table}

\subsubsection{Business Data}
\label{sec:businessdata}

\newcolumntype{e}{>{\hsize=.005\hsize}X}
\newcolumntype{f}{>{\hsize=.045\hsize}X}
\newcolumntype{g}{>{\hsize=.08\hsize}X}
\newcolumntype{h}{>{\hsize=.55\hsize}X}
\newcolumntype{i}{>{\hsize=.25\hsize}X}
\begin{table*}[]
\caption{ESPBench query set (based on~\protect\cite{DBLP:conf/tpctc/HesseRMLKU17})}
\vspace{-1em}
\label{tab:queries}

\begin{tabularx}{\linewidth}{@{}egfhi@{}}
\toprule
\footnotesize \# &
\footnotesize Use Case &
\footnotesize Tested Aspects &
\footnotesize Query Definition &
\footnotesize Description \\ \midrule
\footnotesize 1
&
\footnotesize Check Sensors
&
\footnotesize 1;2;3
&
\fontsize{6}{8}\selectfont \textbf{SELECT} AVG(mf01), MIN(mf01), MAX(mf01), COUNT(mf01) \newline \textbf{FROM} STREAM\_SENSOR
\textbf{TUMBLING WINDOW} 1 \textbf{SECONDS}
&
\fontsize{8}{8}\selectfont Calculate \textit{avg, min, max, count} for the last 1sec for \textit{mf01} for monitoring.
\\
\footnotesize 2
&
\footnotesize Determine Outliers
&
\footnotesize 1;6
&
\fontsize{6}{8}\selectfont
\textbf{SELECT} STOCHASTIC\_OUTLIERS(mf01, mf02),
\newline outlier\_probability
\newline
\textbf{FROM} STREAM\_SENSOR
\newline
\textbf{CUSTOM WINDOW} 500 \textbf{ELEMENTS}  \textbf{WHERE} threshold $>=$ 0.5
&
\fontsize{8}{8}\selectfont
Calculate outliers using Stochastic Outlier Selection~\cite{sos} for combination of \emph{mf01} and \emph{mf02}. Output records that are an outlier with at least 50\% probability.
\\
\footnotesize 3
&
\footnotesize Identify Errors
&
\footnotesize 4
&
\fontsize{6}{8}\selectfont

\textbf{SELECT} * \textbf{FROM} STREAM\_SENSOR \textbf{WHERE} mf01 $>$ 14,963
&
\fontsize{8}{8}\selectfont
Log if sensor value electrical power main phase 1 exceeds limit of 14,963.
\\

\footnotesize 4
&
\footnotesize Check Machine Power
&
\footnotesize 5;7
&
\fontsize{6}{8}\selectfont

\textbf{SELECT} * \textbf{FROM} STREAM\_SENSOR1 \textbf{AS} s1,
STREAM\_SENSOR2 \textbf{AS} s2,  DB\_TABLE\_1 \textbf{AS} t
\textbf{WHERE}
\newline
(s1.M\_ID = t.M\_ID
 \textbf{AND}
s1.mf03 $<$ 8,105
\textbf{AND}
(s1.TS $>$ t.DOWNT\_END
\textbf{OR}
s1.TS $<$ t.DOWNT\_START))
\newline
\textbf{OR}
(s2.M\_ID = t.M\_ID
 \textbf{AND}
s2.mf03 $<$ 8,105
\textbf{AND}
(s2.TS $>$ t.DOWNT\_END
\textbf{OR}
s2.TS $<$ t.DOWNT\_START))
&
\fontsize{8}{8}\selectfont
Log if any machine is in an unscheduled phase of being turned off or in standby (assumption: there is always the next downtime stored in DB\_TABLE\_1).
\\
\footnotesize 5
&
\footnotesize Persist Processing Times for Products
&
\footnotesize
4;7
&
\fontsize{6}{8}\selectfont
\textbf{UPDATE} PRODUCTION\_ORDER\_LINE
\textbf{IF} (STREAM\_TIMES.PT\_IS\_END == 0) \{
\newline \textbf{SET} POL\_START\_TS =
(\textbf{SELECT} TIMESTAMP
\textbf{FROM} STREAM\_TIMES) \}
\newline \textbf{ELSE} \{
\textbf{SET} POL\_END\_TS =
(\textbf{SELECT} TIMESTAMP
\textbf{FROM} STREAM\_TIMES)
\}
\newline \textbf{WHERE} POL\_O\_ID = STREAM\_TIMES.PT\_O\_ID
 \textbf{AND} POL\_OL\_NUMBER =
 \newline STREAM\_TIMES.PT\_OL\_NUMBER
 \textbf{AND} POL\_NUMBER = STREAM\_TIMES.PT\_POL\_NUMBER
&
\fontsize{8}{8}\selectfont
Whenever a product enters or leaves a workplace, log the time to the corresponding DBMS entry (PRODUCTION\_ORDER\_LINE table).
\\  \bottomrule
\end{tabularx}
\vspace{-1em}
\end{table*}

\begin{sloppypar}
The schema of the business data is depicted in Figure~\ref{fig:serd}.
It is based on the data schema of the TPC-C benchmark~\cite{DBLP:conf/sigmod/LeuteneggerD93}, one of the most known DBMS benchmarks that uses structured business data.
Its schema covers core business relations, such as \textit{CUSTOMER} and \textit{ORDER}, that are representative of any manufacturing company.
We simplified the TPC-C table design without impacting query costs.
Inspired by modern business systems, we also added new relations that incorporate industrial manufacturing's domain character.
Specifically, we removed the tables \emph{WAREHOUSE}, \emph{STOCK}, \emph{DISTRICT}, \emph{HISTORY}, and \emph{NEW-ORDER}.
We extend the schema by the tables \emph{PRODUCTION\_ORDER}, \emph{PRODUCTION\_ORDER\_LINE}, and \emph{WORKPLACE}, which are highlighted in green in Figure~\ref{fig:serd}.
By default, to have a representative data size, data is generated with a \textit{scale factor} of three, which would equal a TPC-C setting with three warehouses.
This configuration parameter has an impact on the overall business data size and can be altered for scaling reasons.

The introduced table \emph{WORKPLACE} contains information about scheduled downtimes.
These data allow distinguishing planned downtimes from irregularities that require reactions.
The other two added tables contain information about the production orders, which are linked to the customer orders and  workplaces.
Storing business entities, such as sales or production orders, in a header and an item table is a common concept in Enterprise Resource Planning systems~\cite{DBLP:conf/dasfaa/HesseMSU19,DBLP:conf/sigmod/Plattner09}.
\end{sloppypar}
\vspace{-1em}
\subsection{Benchmark Queries}
\label{subsec:ESPBenchqueries}

This section presents the benchmark queries that the SUT is tasked with.
Moreover, we introduce the design objectives that were taken into account when developing these queries.
\vspace{-1em}
\subsubsection{Relevance of Queries}
\label{subsubsec:querydesignobjectives}
When defining benchmark queries, \textit{relevance} and \textit{simplicity} need special consideration.
As outlined in Table~\ref{tab:designcriteria}, having easily understandable queries is a crucial requirement for, e.g., credibility reasons.
Another aspect influenced by simplicity is the application of the benchmark.
A rather straightforward workload is essential for implementing the queries for other streaming architectures at adequate costs, i.e., with implementation efforts for the queries that take a justifiable amount of time.
This simplicity also benefits the usage of the benchmark.

To ensure the \emph{relevance} of queries, the proximity of the workload to real-world scenarios and the coverage of important stream processing functionalities need to be considered.
We addressed the first aspect by discussing our queries regarding their closeness to real-world use cases with two multi-billion revenue manufacturing companies, which confirmed their applicability.

To guarantee that all essential operations of DSPSs are covered, we first need to identify these.
Our definition of these functionalities is based on the core set of operations for event processing systems presented by Mendes~\cite{mendes2014performance}.
Although this set is defined for event processing systems, it is applicable to stream processing in general~\cite{DBLP:conf/tpctc/HesseRMLKU17}.
To incorporate the benchmark's enterprise character, we extend the original list.
Particularly, we broaden the included term \textit{pattern detection} by altering it to \textit{machine learning} to better represent current requirements on DSPSs.
Furthermore, we add the aspects of transforming data, also included in earlier work by Mendes et al.~\cite{DBLP:conf/tpctc/MendesBM09}, and of combining streaming with historical data.
The latter challenge of integrating stored data is one of the eight requirements of real-time stream processing defined by Stonebraker, \c{C}etintemel, and Zdonik~\cite{DBLP:journals/sigmod/StonebrakerCZ05}.

The resulting list of core operations of DSPSs that the queries of ESPBench need to cover to be in line with the design objective of \textit{relevance} is:
\vspace{-0.5em}
\begin{multicols}{2}
\begin{enumerate}
	\item Windowing
	\item Transformation
	\item Aggregation/Grouping
	\item Filtering (Selection / Projection)
	\item Correlation / Enrichment (Join)
	\item Machine Learning
	\item Combination with Historical Data
\end{enumerate}
\end{multicols}
\vspace{-1em}
\subsubsection{Query Definitions}
\label{subsubsec:querydefinitions}

Table~\ref{tab:queries} contains the query set of ESPBench.
It specifically shows a query id, a brief description of the use case, the tested core functionalities of DSPSs, the exact query definition, and a more detailed explanation of the scenario and motivation.
The queries of ESPBench fully cover the previously presented list of core operations.
ESPBench defines the queries using a syntax inspired by the \emph{Continuous Query Language} (CQL)~\cite{DBLP:journals/vldb/ArasuBW06}.
CQL is based on the popular \emph{Structured Query Language} (SQL) and extends it by incorporating data streams and corresponding constructs for data stream processing.
However, CQL is, unlike SQL, not a language that is supported by a majority of DSPSs.

\vspace{-0.5em}
\paragraph{Query 1 - Check Sensor Status}
The query \emph{Check Sensors} monitors the attribute \textit{mf01}, i.e., the electrical power in main phase one, to allow operators insights into irregularities as soon as possible by providing useful and up-to-date key performance indicators (KPIs).
The query calculates the average, minimum, maximum, and the overall number of sensor values in tumbling windows of one second.
The result records contain these calculated KPIs separated by comma.
\vspace{-0.5em}
\paragraph{Query 2 - Determine Outliers}
Th second query determines outliers based on the input of \emph{mf01} and \emph{mf02}.
Its results give hints on irregularities in the manufacturing equipment.
We employ the \emph{stochastic outlier selection} algorithm~\cite{sos} on count-based tumbling windows with 500 elements.
The query outputs values that are an outlier with a probability of $\geqslant 50\%$ in order to identify possible irregularities.
Structurally, the output is represented as the corresponding input sensor record plus the outlier probability correct to two decimal places, which is separated from the corresponding sensor record by a comma.
\vspace{-0.5em}
\paragraph{Query 3 - Identify Errors}
Query three reports actual errors, which are defined by unusually high power consumption of a machine, i.e., greater than 14,963, the 99.5\%tile in main phase one.
The output of query three is the corresponding input sensor record.
Benchmark runs with input rates of 1K and 10K\,messages/second output about 4 and 40\,errors/second, respectively.
\vspace{-0.5em}
\paragraph{Query 4 - Check Machine Power}
This query checks if the power is unexpectedly low, which is a state requiring actions.
As input, the query gets two structurally identical data streams from two machines and business data.
If any of the machines are in an unplanned phase of being shut-down or on standby, the corresponding record needs to be logged.
This is the case if \textit{mf03} falls below the value of 8,105, the 9\%tile, and there is no downtime planned for the machine.
Planned downtimes are persisted in the DBMS that is part of the SUT.
The table \textit{WORKPLACE} stores the beginning (\textit{WP\_DOWNTIME\_START}) and end (\textit{WP\_DOWNTIME\_END}) of the next scheduled downtime for any machine identifiable by its ID (\textit{WP\_ID}).
This machine or workplace identifier is part of the sensor data as shown in Table~\ref{tab:sensdata}.
\vspace{-0.5em}
\paragraph{Query 5 - Persist Processing Times}
Query five represents another use case where sensor data and historical business data are combined, which highlights the enterprise character of ESPBench.
Particularly, this query stores time data in the DBMS, which is contrary to query four that reads business data.
Having this information allows for data integration, i.e., for connecting sensor data with business data, by using a timestamp-based approach.
Our industry studies revealed that this is a commonly applied technique and thus, incorporating it strengthens the \textit{relevance} of ESPBench.

The DBMS relation \textit{DB\_PRODUCTION\_ORDER\_LINE}, which contains information about the factory's production orders, needs to be updated by query five.
Data input for this query is the \emph{production times stream} with the structure depicted in Table~\ref{tab:stream2_struct}.
The contained data indicates when a product or a part of it entered or left a workplace.
The current timestamp needs to be set in table \textit{PRODUCTION\_ORDER\_LINE}, either in the start or end timestamp column, depending on the incoming sensor record.
\vspace{-0.5em}
\section{ESPBench Validation Setup}
\label{sec:hbvali}

This section gives details on the ESPBench validation, particularly on its concept, the employed DSPSs, and the technical  setup. 
\vspace{-0.5em}
\subsection{Validation Concept}
To validate the concepts and functioning of ESPBench with its tools, we benchmark three state-of-the-art DSPSs and PostgreSQL, the default DBMS of ESPBench.
We developed and published an example implementation of the benchmark queries using the Apache Beam SDK in version 2.16.0 for these measurements.
This abstraction layer is employed in academic as well as real-world scenarios, e.g., at Lyft~\cite{beamatlyft} and Spotify~\cite{beamatspotify}.
Besides, the execution of Apache Beam applications is not only supported by open-source DSPSs~\cite{beamcapa}, but also by commercial closed-source DSPSs, such as IBM Streams~\cite{ibmstreams}, and Google Cloud Dataflow~\cite{clouddataflow}.
This broad application of Apache Beam, both from a user perspective and a DSPS perspective, illustrates its \emph{relevance}.
Nevertheless, it is important to be aware that the level of effort put into Apache Beam support by DSPSs is likely to be reflected in the performance.
The performance penalty traded for the gain in flexibility is analyzed by Hesse et al.~\cite{DBLP:conf/icdcs/HesseMGHU19}.

\vspace{-2em}
\subsection{Data Stream Processing Systems}
\label{subsec:benchmarkedSystems}

The three analyzed DSPSs are \emph{Hazelcast Jet}~\cite{hcjet}, \emph{Apache Flink}~\cite{DBLP:journals/debu/CarboneKEMHT}, and \emph{Apache Spark Streaming}~\cite{DBLP:conf/sosp/ZahariaDLHSS13}, which is an extension to the \emph{Apache Spark}~\cite{DBLP:conf/hotcloud/ZahariaCFSS10} system.
All of these systems are mainly written in a Java Virtual Machine (JVM) language, i.e., Java or Scala.
Apache Spark Streaming processes micro-batches, while the other systems process records tuple-by-tuple.
While Apache Flink and Spark follow a master-worker pattern in their system design, a Hazelcast Jet cluster only contains nodes of the same kind.
All systems are able to provide exactly-once processing guarantees~\cite{jetfaq,DBLP:conf/icpads/HesseL15}.

As Hazelcast Jet is a less popular system compared to Flink and Spark, we introduce it in the following paragraph.
Hazelcast Jet, short Jet, is the stream processing engine of the \emph{Hazelcast} company, headquartered in Silicon Valley.
The streaming engine uses the \emph{Hazelcast distributed in-memory data grid} (IMDG), short Hazelcast.
Next to open-source versions of Hazelcast IMDG and Hazelcast Jet published under the Apache License~\cite{githubhazelcastjet,githubhazelcast}, the commercial versions \emph{Hazelcast IMDG Enterprise} and \emph{Hazelcast IMDG Enterprise HD} are also offered.
The latter one extends Hazelcast IMDG Enterprise by, e.g., a high-density (HD) memory store.
There is also a Hazelcast Jet Enterprise component, which extends Hazelcast Jet by, e.g., security features and a lossless restart functionality~\cite{hazelcastmanual}.

From an architectural perspective, Jet is different from Apache Spark and Apache Flink due to its masterless design.
The oldest node in the cluster represents the de facto leader and manages data responsibilities within the cluster.
Hazelcast organizes the data in shards or partitions, which are distributed equally among the cluster.
It keeps data backups at multiple nodes to prevent data loss in case of a node failure~\cite{johns2015getting}.

\begin{figure}[]
\centering
\includegraphics[width=0.6\columnwidth]{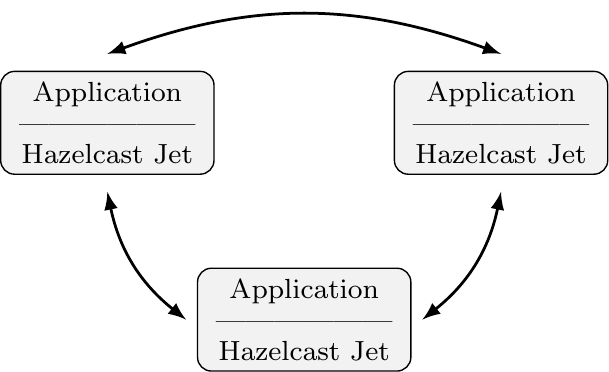}
\caption{Architecture of an embedded Hazelcast Jet deployment with three cluster nodes (based on\protect\cite{hazelcastmanual,johns2015getting})}
\label{fig:hazelcastjetembedded}
\end{figure}

\begin{figure}[]
\centering
\includegraphics[width=0.7\columnwidth]{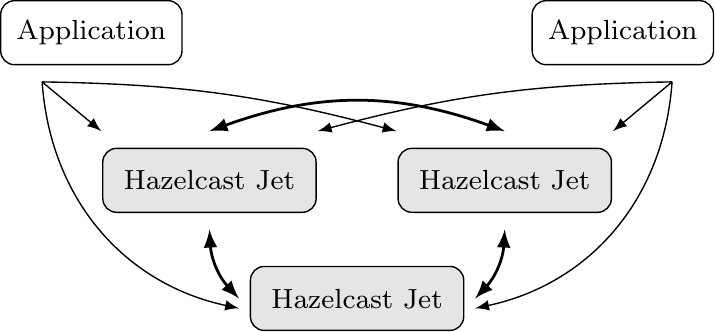}
\caption{Architecture of a client-server Hazelcast Jet deployment with three cluster nodes and two apps (based on\protect\cite{hazelcastmanual,johns2015getting})}
\label{fig:hazelcastjetcs}
\vspace{-1em}
\end{figure}

\begin{sloppypar}
Two deployment modes are supported for both Hazelcast IMDG and Jet: the embedded mode and the client-server mode.
Figures~\ref{fig:hazelcastjetembedded} and \ref{fig:hazelcastjetcs} visualize the embedded and client-server deployments, respectively.
In embedded deployment mode, both the application as well as Jet share a single JVM.
This design has the advantage of low-latency data access due to the tight coupling of application and Jet.
A downside of the embedded mode is its inability to scale the application and data processing engine/data persistence independently~\cite{hazelcastmanual,johns2015getting}.
In contrast, a client-server deployment may create, scale, and manage the cluster independently from any application executed on it.
This setup allows separation of application and cluster but also introduces challenges.
For instance, classpath management of both, application and cluster nodes, requires more attention~\cite{johns2015getting}.
For validating ESPBench, we employ the client-server deployment, which is is more relevant in corporate contexts, as the provided scalability and flexibility aspects are standard requirements for enterprise IT landscapes.
\end{sloppypar}
\vspace{-0.5em}

\subsection{Server Landscape}
\label{subsec:benchmarkSetup}

The validation setup consists of eight virtual machines (VMs), each of which exclusively uses their underlying server.
This allows running the benchmark components independently of each other.

One of these servers serves as the starting point, from which the Ansible script that automates the benchmark execution is started.
The DBMS is deployed on another VM.
Three further nodes build the Apache Kafka cluster and the remaining three VMs contain the DSPS.
Hazelcast Jet is deployed in the client-server mode using all three nodes.
Apache Spark and Apache Flink use two nodes as workers and one node as the master.

The system characteristics of the nodes employed for the experiments are listed in Tables~\ref{tab:system} and~\ref{tab:system2}.
All system configurations are part of the ESPBench repository.
To ensure that Apache Kafka does not become a bottleneck, i.e., to make sure that the SUT is benchmarked as intended, we employ data rates that can provably be handled by Apache Kafka in the described setup.
The study presented by Hesse et al.~\cite{DBLP:journals/corr/kafka} indicates that Apache Kafka can easily handle 1K as well as 10K\,messages/second.
Our conducted industry studies revealed that these are satisfying ingestion rates for the majority of scenarios in the industrial manufacturing sector.
Besides, the work shows that input rates in such a range with the used input data characteristics do not saturate the network capacities.

\begin{table}[!htb]
\caption{System characteristics of the Apache Kafka brokers}
\vspace{-1em}
\label{tab:system}
\small
\begin{tabularx}{\columnwidth}{@{}lX@{}}
\toprule
Characteristic & Value \\ \midrule
Operating System               &  Ubuntu 18.04 LTS    \\
    CPU           &   Intel(R) Xeon(R) CPU X7560 @ 2.27GHz, 8 cores (2x), \newline Intel(R) Xeon(R) CPU E5-2697 v3 @ 2.60GHz, 8 cores (1x)    \\
    RAM           &   57GB (2x), 32GB (1x)    \\
    Network      & 10Gbit via Fujitsu PRIMERGY BX900 S1 \\
    Disk & 13 Seagate ST320004CLAR2000 in RAID 6, access via Fibre Channel with  8Gbit/s\\
    Hypervisor & VMware ESXi 6.7.0\\
    Apache Kafka Version           &   2.3.0    \\
    Java Version & OpenJDK 1.8.0\_222 \\ \bottomrule
\end{tabularx}
\vspace{-1em}
\end{table}

\begin{table}[!htb]
\caption{System characteristics of the SUT nodes}
\vspace{-1em}
\label{tab:system2}
\small
\begin{tabularx}{\columnwidth}{@{}lX@{}}
\toprule
Characteristic & Value \\ \midrule
Operating System               &  Ubuntu 18.04 LTS    \\
    CPU           &   Intel(R) Xeon(R) CPU E5450 @ 3.00GHz, 8 cores  \\
    RAM           &   57GB    \\
    Network      & 10Gbit via Fujitsu PRIMERGY BX900 S1 \\
    Disk & 13 Seagate ST320004CLAR2000 in RAID 6, access via Fibre Channel with  8Gbit/s\\
    Hypervisor & VMware ESXi 6.7.0\\
    Apache Flink Version           &   1.8.2    \\
    Hazelcast Jet Version           &   3.0    \\
    Apache Spark Version           &   2.4.4    \\
    PostgreSQL Version           &   9.6.12    \\
    Scala Version           &    2.12.8   \\
    Java Version & OpenJDK 1.8.0\_222 \\ \bottomrule
\end{tabularx}
\vspace{-1.5em}
\end{table}

\section{Experimental Evaluation}
\label{sec:evaluation}

This section presents the results of the experimental evaluation of ESPBench, which are also made available online\footnote{\url{https://github.com/guenter-hesse/ESPBenchExperiments}}.
After giving an overview of the measurements, we analyze each query in detail.
These individual analyses include a view on the system loads, which ESPBench collects every ten seconds in the applied settings.
The system load gives an overview over the CPU and I/O utilization of a server, i.e., also reflecting performance limits regarding disk writes.
It is defined as the number of processes demanding CPU time, specifically processes that are ready to run or waiting for disk I/O.
The included figures visualize one-minute-averages of this system load KPI.
As we are using servers with an eight-core CPU, described in Section~\ref{subsec:benchmarkSetup}, it is desirable that no node exceeds a system load of eight to do not over-utilize a machine~\cite{systemload}.

\subsection{Result Overview}
\label{subsubsec:resultoverview}

\begin{table*}[]
\caption{Latency result overview of the experimental analysis conducted as part of the validation of ESPBench}
\label{tab:overallresults}
\centering
\begin{tabular}{lrcrrrr}
\hline
\multicolumn{1}{c}{Query} &
  \multicolumn{1}{c}{
  \begin{tabular}[c]{@{}c@{}}Input Rate in\\messages/second\end{tabular}
  }
  &
  System &
  \multicolumn{1}{c}{90\%tile in s} &
  \multicolumn{1}{c}{Min in s} &
  \multicolumn{1}{c}{Max in s} &
  \multicolumn{1}{c}{Mean in s} \\ \hline
1 - Check Sensors
& 1,000
& \cellcolor[HTML]{d4f4d4} Apache Flink
& \cellcolor[HTML]{d4f4d4} 10.659
& \cellcolor[HTML]{d4f4d4}0.049
& \cellcolor[HTML]{d4f4d4}18.591
& \cellcolor[HTML]{d4f4d4}4.269  \\

&
& \cellcolor[HTML]{e4e5f9} Hazelcast Jet
& \cellcolor[HTML]{e4e5f9}0.024
& \cellcolor[HTML]{e4e5f9}0.009
& \cellcolor[HTML]{e4e5f9}0.691
& \cellcolor[HTML]{e4e5f9}0.020 \\

&
& \cellcolor[HTML]{fbe5ca} Apache Spark Streaming
& \cellcolor[HTML]{fbe5ca}n/a
& \cellcolor[HTML]{fbe5ca}n/a
& \cellcolor[HTML]{fbe5ca}n/a
& \cellcolor[HTML]{fbe5ca}n/a  \\  \cmidrule{2-7} %\hhline{~------}

& 10,000
&  \cellcolor[HTML]{d4f4d4} Apache Flink
&   \cellcolor[HTML]{d4f4d4} 16.492
&  \cellcolor[HTML]{d4f4d4} 0.048
&  \cellcolor[HTML]{d4f4d4} 33.423
&  \cellcolor[HTML]{d4f4d4} 5.767  \\

&
& \cellcolor[HTML]{e4e5f9} Hazelcast Jet
&  \cellcolor[HTML]{e4e5f9}0.036
& \cellcolor[HTML]{e4e5f9}0.012
& \cellcolor[HTML]{e4e5f9}1.030
& \cellcolor[HTML]{e4e5f9}0.029 \\

&
& \cellcolor[HTML]{fbe5ca} Apache Spark Streaming
& \cellcolor[HTML]{fbe5ca}n/a
&\cellcolor[HTML]{fbe5ca} n/a
& \cellcolor[HTML]{fbe5ca}n/a
& \cellcolor[HTML]{fbe5ca}n/a   \\ \cmidrule{2-7} %\hhline{~------}
2 - Determine Outliers
& 1,000
& \cellcolor[HTML]{d4f4d4} Apache Flink
& \cellcolor[HTML]{d4f4d4} 615.078
& \cellcolor[HTML]{d4f4d4}9.352
& \cellcolor[HTML]{d4f4d4}676.535
& \cellcolor[HTML]{d4f4d4}358.076  \\

&
& \cellcolor[HTML]{e4e5f9} Hazelcast Jet
&  \cellcolor[HTML]{e4e5f9}533.177
& \cellcolor[HTML]{e4e5f9}5.353
& \cellcolor[HTML]{e4e5f9}590.170
& \cellcolor[HTML]{e4e5f9}304.689  \\

&
& \cellcolor[HTML]{fbe5ca} Apache Spark Streaming
& \cellcolor[HTML]{fbe5ca}n/a
& \cellcolor[HTML]{fbe5ca}n/a
& \cellcolor[HTML]{fbe5ca}n/a
& \cellcolor[HTML]{fbe5ca}n/a   \\ \cmidrule{2-7} %\hhline{~------}

& 10,000
& \cellcolor[HTML]{d4f4d4} Apache Flink
&  \cellcolor[HTML]{d4f4d4}8,175.784
&\cellcolor[HTML]{d4f4d4} 40.446
& \cellcolor[HTML]{d4f4d4}9,147.738
& \cellcolor[HTML]{d4f4d4}4,599.666  \\

&
&\cellcolor[HTML]{e4e5f9} Hazelcast Jet
& \cellcolor[HTML]{e4e5f9} 7,425.443
& \cellcolor[HTML]{e4e5f9}24.564
& \cellcolor[HTML]{e4e5f9}8,282.022
& \cellcolor[HTML]{e4e5f9}4,140.149  \\

&
& \cellcolor[HTML]{fbe5ca} Apache Spark Streaming
& \cellcolor[HTML]{fbe5ca}n/a
& \cellcolor[HTML]{fbe5ca}n/a
& \cellcolor[HTML]{fbe5ca}n/a
& \cellcolor[HTML]{fbe5ca}n/a   \\ \cmidrule{2-7} %\hhline{~------}
3 - Identify Errors
& 1,000
& \cellcolor[HTML]{d4f4d4} Apache Flink
& \cellcolor[HTML]{d4f4d4}0.011
& \cellcolor[HTML]{d4f4d4}0.001
& \cellcolor[HTML]{d4f4d4}0.045
& \cellcolor[HTML]{d4f4d4}0.005  \\

&
& \cellcolor[HTML]{e4e5f9} Hazelcast Jet
& \cellcolor[HTML]{e4e5f9} 0.021
& \cellcolor[HTML]{e4e5f9}0.004
& \cellcolor[HTML]{e4e5f9}0.158
& \cellcolor[HTML]{e4e5f9}0.017  \\

&
& \cellcolor[HTML]{fbe5ca} Apache Spark Streaming
& \cellcolor[HTML]{fbe5ca}0.534
& \cellcolor[HTML]{fbe5ca}0.121
& \cellcolor[HTML]{fbe5ca}1.248
& \cellcolor[HTML]{fbe5ca}0.387  \\ \cmidrule{2-7} %\hhline{~------}

& 10,000
& \cellcolor[HTML]{d4f4d4} Apache Flink
& \cellcolor[HTML]{d4f4d4} 14.979
& \cellcolor[HTML]{d4f4d4}0.002
& \cellcolor[HTML]{d4f4d4}19.058
& \cellcolor[HTML]{d4f4d4}4.581  \\

&
& \cellcolor[HTML]{e4e5f9} Hazelcast Jet
& \cellcolor[HTML]{e4e5f9} 0.016
& \cellcolor[HTML]{e4e5f9}0.005
& \cellcolor[HTML]{e4e5f9}0.795
& \cellcolor[HTML]{e4e5f9}0.014  \\

&
& \cellcolor[HTML]{fbe5ca} Apache Spark Streaming
&  \cellcolor[HTML]{fbe5ca}1.557
& \cellcolor[HTML]{fbe5ca}0.137
& \cellcolor[HTML]{fbe5ca}5.380
& \cellcolor[HTML]{fbe5ca}0.780  \\ \cmidrule{2-7} %\hhline{~------}
4 - Check Machine Power
& 1,000
& \cellcolor[HTML]{d4f4d4} Apache Flink
& \cellcolor[HTML]{d4f4d4}0.717
& \cellcolor[HTML]{d4f4d4}0.003
& \cellcolor[HTML]{d4f4d4}2.792
& \cellcolor[HTML]{d4f4d4}0.251  \\

&
& \cellcolor[HTML]{e4e5f9} Hazelcast Jet
& \cellcolor[HTML]{e4e5f9} 0.371
& \cellcolor[HTML]{e4e5f9}0.006
& \cellcolor[HTML]{e4e5f9}4.082
& \cellcolor[HTML]{e4e5f9}0.195  \\

&
& \cellcolor[HTML]{fbe5ca} Apache Spark Streaming
&  \cellcolor[HTML]{fbe5ca}1.008
& \cellcolor[HTML]{fbe5ca}0.141
& \cellcolor[HTML]{fbe5ca}1.966
& \cellcolor[HTML]{fbe5ca}0.644  \\ \cmidrule{2-7} %\hhline{~------}

& 10,000
& \cellcolor[HTML]{d4f4d4} Apache Flink
& \cellcolor[HTML]{d4f4d4} 470.689
& \cellcolor[HTML]{d4f4d4}1.936
& \cellcolor[HTML]{d4f4d4}517.291
& \cellcolor[HTML]{d4f4d4}275.096  \\

&
& \cellcolor[HTML]{e4e5f9} Hazelcast Jet
& \cellcolor[HTML]{e4e5f9} 87.299
& \cellcolor[HTML]{e4e5f9}6.008
& \cellcolor[HTML]{e4e5f9}94.599
& \cellcolor[HTML]{e4e5f9}56.236  \\

&
& \cellcolor[HTML]{fbe5ca} Apache Spark Streaming
&  \cellcolor[HTML]{fbe5ca}303.432
& \cellcolor[HTML]{fbe5ca}4.255
& \cellcolor[HTML]{fbe5ca}325.951
& \cellcolor[HTML]{fbe5ca}188.158  \\ \cmidrule{2-7} %\hhline{~------}
5 - Persist Processing Times
& 1,000
& \cellcolor[HTML]{d4f4d4} Apache Flink
& \cellcolor[HTML]{d4f4d4} 106.892
& \cellcolor[HTML]{d4f4d4}0.506
& \cellcolor[HTML]{d4f4d4}114.750
& \cellcolor[HTML]{d4f4d4}65.261  \\
{  }{  }{  }{  }{  }for Products
&
& \cellcolor[HTML]{e4e5f9} Hazelcast Jet
& \cellcolor[HTML]{e4e5f9} 88.006
& \cellcolor[HTML]{e4e5f9}1.823
& \cellcolor[HTML]{e4e5f9}96.316
& \cellcolor[HTML]{e4e5f9}51.278  \\

&
& \cellcolor[HTML]{fbe5ca} Apache Spark Streaming
&  \cellcolor[HTML]{fbe5ca}102.736
& \cellcolor[HTML]{fbe5ca}0.803
& \cellcolor[HTML]{fbe5ca}112.815
& \cellcolor[HTML]{fbe5ca}61.820  \\ \cmidrule{2-7} %\hhline{~------}

& 10,000
& \cellcolor[HTML]{d4f4d4} Apache Flink
& \cellcolor[HTML]{d4f4d4} 2,028.137
& \cellcolor[HTML]{d4f4d4}2.274
& \cellcolor[HTML]{d4f4d4}2,211.899
& \cellcolor[HTML]{d4f4d4}1,136.910  \\

&
& \cellcolor[HTML]{e4e5f9} Hazelcast Jet
& \cellcolor[HTML]{e4e5f9} 2,129.287
& \cellcolor[HTML]{e4e5f9}6.202
& \cellcolor[HTML]{e4e5f9}2,345.790
& \cellcolor[HTML]{e4e5f9}1,170.944  \\

&
& \cellcolor[HTML]{fbe5ca} Apache Spark Streaming
&  \cellcolor[HTML]{fbe5ca}1,863.259
& \cellcolor[HTML]{fbe5ca}1.941
& \cellcolor[HTML]{fbe5ca}2,061.002
& \cellcolor[HTML]{fbe5ca}1,041.930  \\ \hline
\end{tabular}
\end{table*}

\begin{sloppypar}
Table~\ref{tab:overallresults} shows a summary of the results, i.e., the 90\%tile, minimum, maximum, and mean latencies for the different ESPBench queries and benchmark settings.
Each query was executed three times on every system and with each data input rate, which is sufficient due to the low variance of latency results.
We benchmarked data input rates of 1K and 10K\,messages/second.
A single benchmark run lasts five minutes.
Overall, the latency results are diverse with Jet often performing best with respect to the 90\%tile and mean values.
Next to Table~\ref{tab:overallresults}, Figure~\ref{fig:graphs} visualizes single latencies and system loads for selected queries and settings.
\end{sloppypar}

\subsection{Query 1 - Check Sensor Status}
\label{subsubsec:q1}

The results shown in Table~\ref{tab:overallresults} reveal that Jet performed significantly better than Apache Flink.
Although Apache Flink's minimum latency is relatively low (49\,ms), there are remarkably higher latencies, the maximum being above 18\,s and the 90\%tile being at more than 10\,s.
In contrast, Jet's worst response time is about 700\,ms.

Apache Spark's response times cannot be compared as the query results differ from the expected results.
This is the case although we use the same application developed using the Apache Beam SDK for all systems.
This valuable finding highlights the importance of having a query result validation for performance benchmarks as published with ESPBench.
An explanatory hypothesis for the false results is related to the data architecture of Apache Spark Streaming, i.e., the distinguishing use of micro-batches.
Through batching mechanisms, windows might look different.
If a micro-batch represents the finest granularity and cannot be split, it could be left out of a window even though most of the contained records semantically belong to the window, depending on the window semantics of the DSPS.
The work by Botan et al.~\cite{DBLP:journals/pvldb/BotanDDHMT10} studies the heterogeneity in window semantics that exists in DSPSs.

\begin{figure*}[]
\centering
	\begin{subfigure}[t]{0.315\textwidth}
	\includegraphics[width=\textwidth]{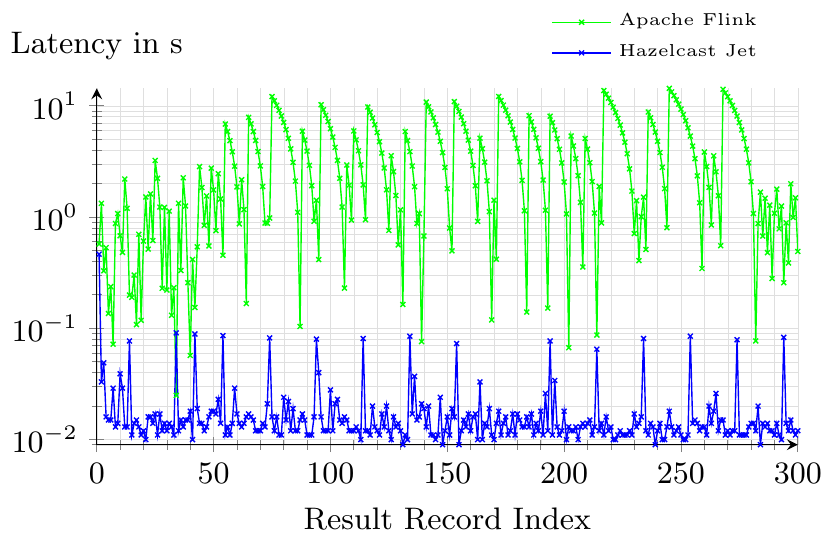}
	\caption{Latencies query 1 for a data input rate of 1,000\,messages/second}
	\label{fig:q1}
	\end{subfigure}
	\quad
	\begin{subfigure}[t]{0.315\textwidth}
		\includegraphics[width=\textwidth]{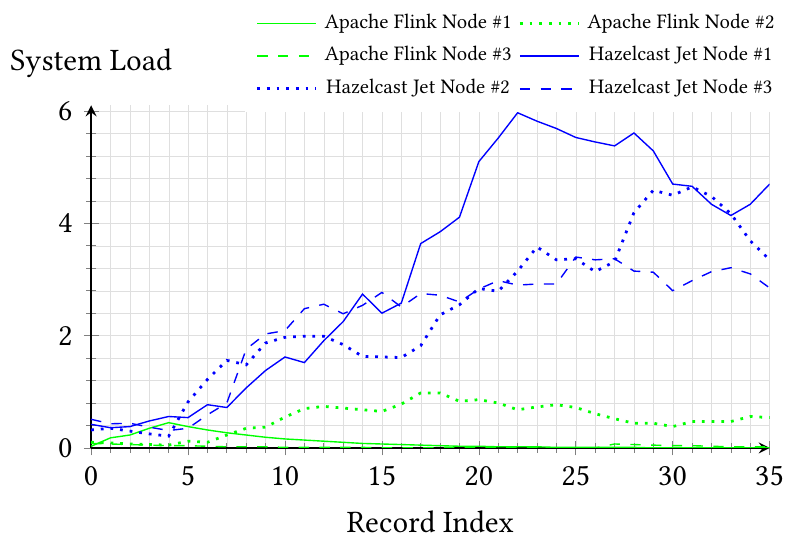}
		\caption[One-minute-average of short-term system load for query 1 for a data input rate of 1,000\,messages/second]{One-minute-average of short-term system load for query 1 for a data input rate of 1,000\,messages/second}
		\label{fig:q1load}
	\end{subfigure}
	\quad
	\begin{subfigure}[t]{0.315\textwidth}
	\includegraphics[width=\textwidth]{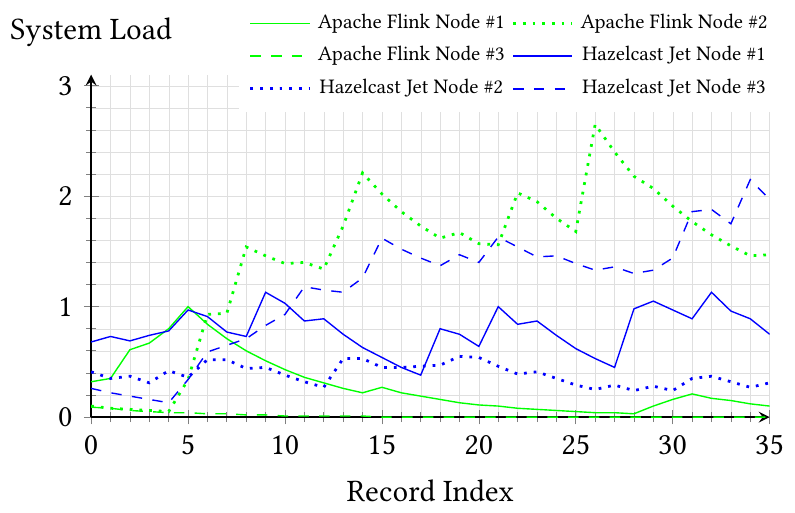}
	\caption{One-minute-average of short-term system load for query 2 for a data input rate of 1,000\,messages/second}
	\label{fig:q2load}
	\end{subfigure}

	\begin{subfigure}[t]{0.315\textwidth}
		\includegraphics[width=\textwidth]{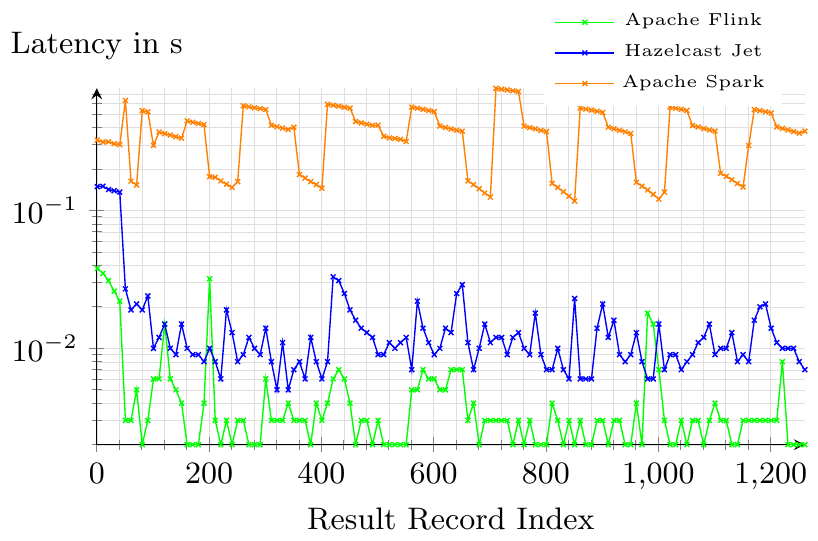}
		\caption{Latencies query 3 for a data input rate of 1,000\,messages/second}
		\label{fig:q3}
	\end{subfigure}
	\quad
	\begin{subfigure}[t]{0.315\textwidth}
		\includegraphics[width=\textwidth]{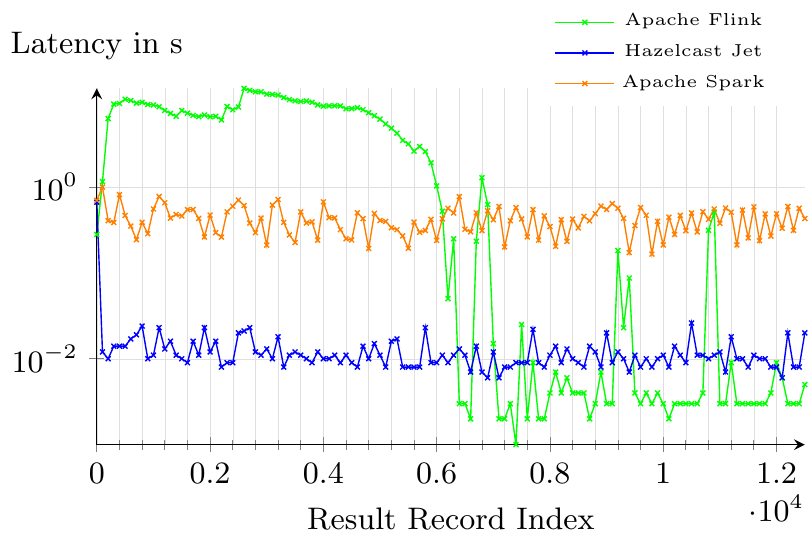}
		\caption{Latencies query 3 for a data input rate of 10,000\,messages/second}
		\label{fig:q3_10k}
	\end{subfigure}
	\quad
	\begin{subfigure}[t]{0.315\textwidth}
		\includegraphics[width=\textwidth]{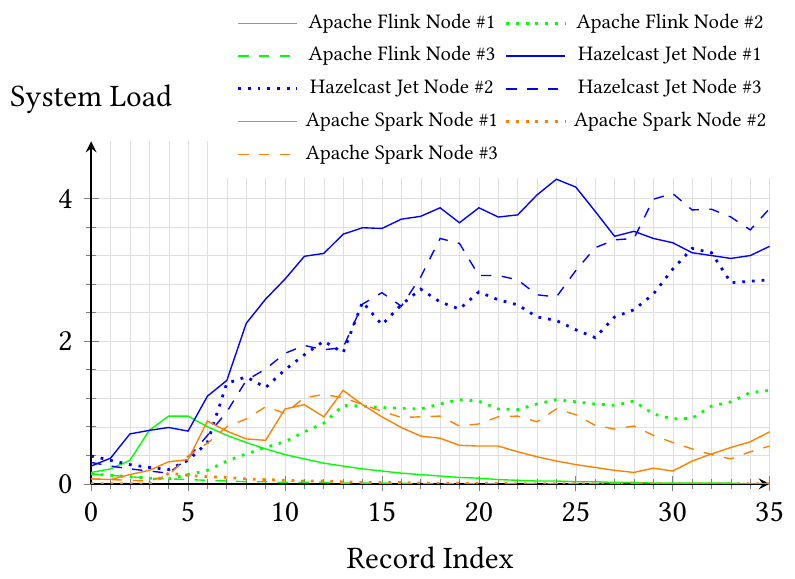}
		\caption{One-minute-average of short-term system load for query 3 for a data input rate of 10,000\,messages/second}
		\label{fig:q3load}
	\end{subfigure}

	\begin{subfigure}[t]{0.315\textwidth}
		\includegraphics[width=\textwidth]{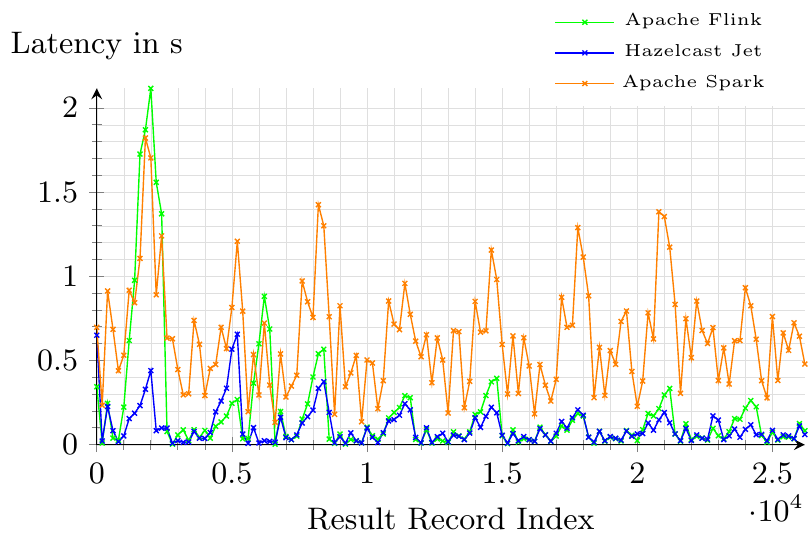}
		\caption{Latencies query 4 for a data input rate of 1,000\,messages/second}
		\label{fig:q4}
	\end{subfigure}
	\quad
	\begin{subfigure}[t]{0.32\textwidth}
		\includegraphics[width=\textwidth]{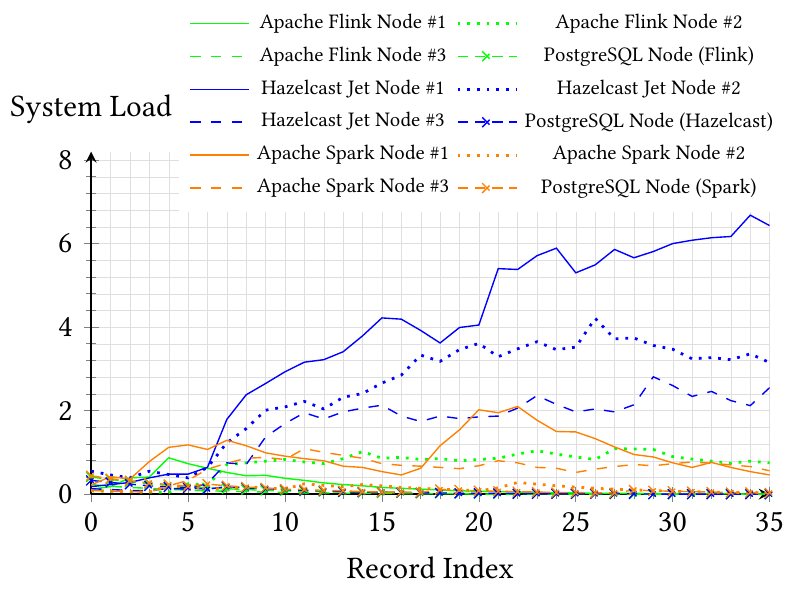}
		\caption{One-minute-average of short-term system load for query 4 for a data input rate of 1,000\,messages/second}
		\label{fig:q4load}
	\end{subfigure}
	\quad
	\begin{subfigure}[t]{0.32\textwidth}
		\includegraphics[width=\textwidth]{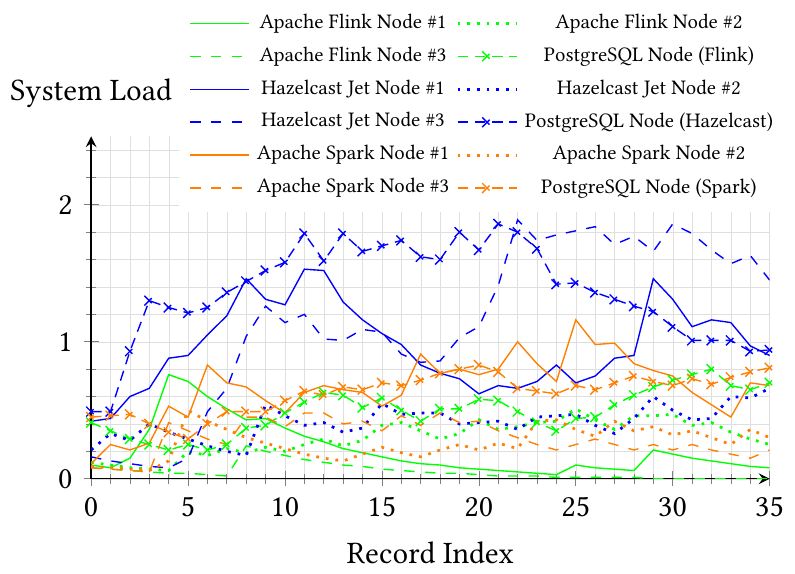}
		\caption{One-minute-average of short-term system load for query 5 for a data input rate of 1,000\,messages/second}
		\label{fig:q5load}
	\end{subfigure}
%\vspace{-1em}
\caption{Selected latency and system load graphs of the experiments conducted as part of the validation of ESPBench}
\vspace{-1em}
\label{fig:graphs}
\end{figure*}

Figure~\ref{fig:q1} visualizes the result latencies of query one for the input rate of 1K\,messages/second.
It becomes visible that there are upward outliers for both systems, while the overall latencies on Apache Flink are significantly higher as identified before.
After the Apache Flink latency reaches a new local maximum, the latencies slowly decline step-by-step.
This behavior is different from the latency development that we can observe for Jet runs, where the latency, after facing an upward outlier, immediately jumps back to normal, i.e., the value range to which most latencies belong.

The gathered system characteristics reveal that Hazelcast utilizes the cluster more. 
Figure~\ref{fig:q1load} shows the system load for the SUT nodes while executing the first query.
Two major differences between Apache Flink and Hazelcast Jet become visible.
Firstly, all Hazelcast Jet nodes show a higher system load than the Apache Flink node with the highest utilization.
Secondly, while Hazelcast Jet utilizes all three nodes, which results in a load between approximately two and six, there is only one Apache Flink node that shows a utilization close to one.
The better system utilization is likely to be a reason for the lower latencies that are associated with Hazelcast Jet for query one compared to Apache Flink, cf. Table~\ref{tab:overallresults}.
%\vspace{-1em}
\subsection{Query 2 - Determine Outliers}
\label{subsubsec:q2}

\begin{sloppypar}
	The latencies of Table~\ref{tab:overallresults} show significantly higher values for query two than for query one, with Hazelcast Jet slightly outperforming Apache Flink. 
Executing the query implementation on the Apache Spark Streaming system did not reveal any latencies, but another valuable finding.
Specifically, the system throws an exception when submitting the Apache Beam program: \textit{java.lang.IllegalStateException: No TransformEvaluator registered for UNBOUNDED transform View.CreatePCollectionView}.
This observation indicates that the DSPS exchangeability of Apache Beam applications is limited.
\end{sloppypar}

The latencies of both systems show a steadily growing latency~\footnote{\url{https://github.com/guenter-hesse/ESPBenchExperiments}, accessed: 2021-02-01}, which indicates queuing in the system.
That means the outlier detection cannot be performed fast enough for the configured input rate.
This queuing does not become visible when looking at the overall numbers shown in Table~\ref{tab:overallresults}, which highlights the importance of the ESPBench feature to output single record latencies.

Figure~\ref{fig:q2load} visualizes the system loads for both systems.
The load is generally lower and both systems create more similar system loads compared to query one.
These measurements suggest that the implementation of the stochastic outlier selection has room for improvement regarding its performance, e.g., by parallelizing it.

\subsection{Query 3 - Identify Errors}
\label{subsubsec:q3}

Figure~\ref{fig:q3} shows the latencies for the input of 1K\,messages/second, specifically every tenth latency for readability reasons.
It illustrates that there are ups and downs that stay in system-specific ranges, Apache Flink having the lowest latencies, followed by Hazelcast Jet and Apache Spark Streaming.
Apache Spark Streaming further shows a pattern-like trend, which can be due to the use of micro-batches.

Figure~\ref{fig:q3_10k} shows every 100\textsuperscript{th} latency for 10K\,messages/second.
Hazelcast Jet and Apache Spark Streaming again show ups and downs within a comparatively small range.
The previously observable pattern for Apache Spark is not present anymore.
Apache Flink presents a different behavior.
For about half of the five-minute benchmark run, the latencies are on a relatively high level.
After this period, the latencies drop and stay on this new level, with relatively high swings though.
This unique progress can be observed in all Apache Flink runs for this input rate and query.
It again highlights the importance of having and studying single latencies to get full insights.

Figure~\ref{fig:q3load} visualizes the system load of query three for the input rate of 10K\,messages/second.
The utilization of the Jet runs is similar to the one observed in Figure~\ref{fig:q1load}, i.e., a higher utilization on all three nodes.
Moreover, the utilization of each node is greater than the highest system loads of both, Apache Flink and Apache Spark Streaming runs.

\vspace{-0.5em}
\subsection{Query 4 - Check Machine Power}
\label{subsubsec:q4}

Figure~\ref{fig:q4} visualizes the latencies for a run with an input rate of 1K\,messages/second.
Relatively small peaks can be identified for all systems, which might be caused by garbage collection runs.
While peaks for Apache Spark are comparatively high and constant throughout the benchmark run, the peaks for Apache Flink and Jet are bigger at the beginning with a decreasing trend.
For the input rate of 10K\,messages/second, steadily increasing latencies can be observed, similar to the latency trend for query two.
It again indicates that records were queuing up, i.e., the SUTs cannot handle the higher data input rate in a sustainable fashion.

Figure~\ref{fig:q4load} presents the system loads for query four.
Contrary to the other system load charts, this figure includes the DBMS node as the database is incorporated in the queries four and five.
Figure~\ref{fig:q4load} draws a similar picture as Figure~\ref{fig:q3load} with the system loads for query three, with the difference of having higher loads for Jet and Apache Spark.
Especially node \#1 of the Jet cluster shows the overall highest measured load of about 6.5 for a short period.
The PostgreSQL node shows a very low system load of less than one for all settings.

\vspace{-0.5em}
\subsection{Query 5 - Persist Processing Times}
\label{subsubsec:q5}

The evaluation shows steadily increasing latencies as for query two.
It again indicates that the SUT cannot handle the load properly.
The gathered results reveal that the DBMS is the bottleneck for this write-heavy query with about 300K required updates for the five minute runs with 1K\,messages/second as data input rate. 

Figure~\ref{fig:q5load} visualizes the system loads with an overall maximum below two.
Hazelcast Jet again created the highest system loads compared to the other two DSPSs.
The loads on the PostgreSQL VM are higher than for the previous query, with Hazelcast Jet causing the highest system load on the server where the DBMS is installed.

\vspace{-0.5em}

\section{Threats to Validity}
\label{subsec:summarytdv}

One threat to validity is related to the applicability of results produced by ESPBench to domains different from manufacturing.
The workload of ESPBench belongs to the manufacturing domain and is validated with companies from this industry sector.
Depending on the target domain, the employed combination of data and queries might have a varying value.
However, as the queries cover all core functionalities of DSPSs, results of ESPBench give at least a general hint on a system's performance. 

Another threat to validity is the limited number of metrics employed by ESPBench.
However, it benefits the simplicity of the benchmark.
Regarding the results of the experimental evaluation, the amount of three executions for each benchmark run is another threat to validity, even though variances between runs are low. 

\vspace{-0.5em}

\section{Lessons Learned}
\label{sec:lessonslearned}

We learned three main lessons from developing and evaluating ESPBench, one of them being the importance of result validation, which unfortunately lacks proper tool support in existing benchmarks.
We highlight this importance by pointing to differing results for the same application executed on different DSPSs.
Depending on the scenario, correctness might be more or less important to users of DSPSs.
However, it is crucial to be aware of a system's behavior.

 \begin{sloppypar}
Secondly, the portability of Apache Beam applications is not always given, i.e., it is not guaranteed that the paradigm 'write once, execute anywhere' holds.
In particular, we logged an exception when running the application for query two with Apache Spark after successfully executing it on Apache Flink and Hazelcast Jet.
 \end{sloppypar}

Thirdly, we learned that having single response times is of paramount importance. 
Aggregated KPIs often do not allow to fully understand the system's behavior or are even misleading.
That is the case, e.g., when latencies are steadily growing, i.e., a system cannot handle the load and queues incoming records.
It can be falsely assumed that the mean latency determined after a limited benchmarking period is the one that can be expected in a production deployment, i.e., when the application is running permanently.

\newcolumntype{b}{X}
\newcolumntype{s}{>{\hsize=1.5\hsize}X}
\begin{table*}[!htb]
\caption{Comparison of data stream processing benchmarks}
\vspace{-1em}
\label{tab:rwcomp}
\begin{tabularx}{\linewidth}{@{}aXXXXXX@{}}
\toprule
                                                     & Linear Road~\cite{DBLP:conf/vldb/ArasuCGMMRST04}                                  & StreamBench~\cite{DBLP:journals/concurrency/ShuklaCS17}       & RIoTBench~\cite{DBLP:journals/concurrency/ShuklaCS17}         & 
                                                     YSB~\cite{DBLP:conf/ipps/ChintapalliDEFG16} &
                                                     OSPBench~\cite{DBLP:journals/tpds/DongenP20} &
                                                     \textbf{ESPBench}                             \\ \midrule
\multicolumn{1}{l|}{Benchmark type}                  & application                                  & micro             & mixed             & 
application &
application &
application                          \\
\multicolumn{1}{l|}{Historical data}                 & briefly; one file with historical tolls & -                 & -                 &
briefly; key-value store  with advt. data &
- &
 comprehensively; business data, based on TPC-C \\
\multicolumn{1}{l|}{DSPS functionalities}            & partially covered                            & partially covered & partially covered & 
partially covered &
partially covered &
fully covered                        \\
\multicolumn{1}{l|}{Data sender}                     & stub provided                           & -                 & -                 & 
yes &
yes &
yes                             \\
\multicolumn{1}{l|}{Result validator}                & yes                                     & -                 & -                 & 
yes &
- &
yes                             \\
\multicolumn{1}{l|}{Automation}                      & -                                            & -                 & -                 & 
yes &
partially &
yes                             \\
\multicolumn{1}{l|}{Query implemen-} & -                                            & -                 & yes               & 
yes &
yes &
yes                                  \\ 
\multicolumn{1}{l|}{tations published} &                                           &                &                & 
&
                                  \\ \bottomrule
\end{tabularx}
\vspace{-1em}
\end{table*}

\vspace{-0.5em}
\section{Related Work}
\label{sec:relatedwork}

Though there is an abundance of benchmarks for DBMSs, only a few focus on stream processing architectures.
 \emph{Linear Road} by Arasu et al.~\cite{DBLP:conf/vldb/ArasuCGMMRST04} is one of the most popular benchmarks for DSPSs.
It includes a toolkit comprising a data generator, a result validator, and a data sender stub, i.e., an incomplete implementation of a data sender that is designed to be completed by a benchmark user.
Providing only a stub introduces the danger of different implementations among benchmark users, resulting in less comparable  benchmark results.
Furthermore, the benchmark creators cannot assure that the data sender does not become a bottleneck while benchmarking.
An additional challenge of the Linear Road benchmark is the lack of supporting tools for automation.
The scenario of Linear Road is a variable tolling system for a metropolitan area with multiple expressways.
The accumulated tolls vary and depend on the city’s traffic situation.
Linear Road defines four queries.
However, in the implementations described in the benchmark paper, one of them is skipped due to complexity~\cite{DBLP:conf/vldb/ArasuCGMMRST04}.
Besides streaming data, historical data is sparsely incorporated as tolling history for some queries.

\emph{StreamBench}~\cite{DBLP:journals/concurrency/ShuklaCS17} is a micro benchmark aimed at distributed DSPSs.
It defines a group of seven queries, which contains queries with a single as well as with multiple computational steps.
The employed queries further differ regarding their requirements on keeping state.
Although the queries cover a variety of functionalities, typical streaming operations, such as window functions, are not incorporated.
StreamBench uses two real-world data sets as seeds for data generation. 
Contrary to Linear Road and similarly to ESPBench, StreamBench employs Apache Kafka for decoupling data generation and consumption.
However, a benchmark tool for data ingestion is neither published nor described by the authors of StreamBench.
StreamBench uses multiple result metrics, including latency and throughput.
Additionally, StreamBench presents: a durability index (uptime), a throughput penalty factor (assessing throughput change in a node failure scenario), and a latency penalty factor (assessing latency change in a node failure scenario).
The benchmark does not offer a tool for query result validation.

\emph{RIoTBench}~\cite{DBLP:journals/concurrency/ShuklaCS17} defines micro benchmark as well as application benchmark use cases.
These cover Extract, Transform, and Load (ETL) processes, statistics generation, model training, and predictive analytics scenarios.
RIoTBench uses scaled real-world data sets from Internet of Things (IoT) domains, comprising smart city, smart energy, and health.
Neither a data sender tool nor an application for query result validation is provided by the benchmark.
RIoTBench measures latency, throughput, CPU, and memory utilization, as well as \emph{jitter}.
The latter expresses the difference between the expected and actual output rate during a certain period.

\textit{Yahoo!} published a DSPS benchmark, known as \textit{Yahoo! Streaming Benchmark} (YSB), in 2015~\cite{ysb,DBLP:conf/ipps/ChintapalliDEFG16}. 
YSB distinguishes it from the other benchmarks as it originates from the industry.
However, it contains only a single query in the domain of advertisement.
Van Dongen and Van den Poel~\cite{DBLP:journals/tpds/DongenP20} contributed measurements of the relatively new client library  \textit{Kafka Streams}~\cite{kafkastreams} using their own benchmark.  
However, the authors do not incorporate historical data in the proposed \textit{Open Stream Processing Benchmark} (OSPBench).

Table~\ref{tab:rwcomp} compares the five benchmarks with ESPBench.
In summary, we see the need for a new stream processing benchmark for several reasons.
First, historical data is not or only barely taken into account in all major DSPS benchmarks.
We believe that this is a crucial aspect for enterprises, because, to achieve the greatest added value, streaming data needs to be combined with business data.
Second, the majority of current stream processing benchmarks lack satisfying tool support, e.g., for result validation or data ingestion.
Almost none of the benchmarks supports full automation, which is essential for the simplicity of a stream processing benchmark.
This absence of proper tooling complicates the application of these benchmarks and retrieving objective and credible results.
Finally, the workloads of existing benchmarks fail to cover the core DSPS functionalities, which leads to limited meaningfulness of results.

\vspace{-1em}
\section{Conclusion and Future Work}
\label{sec:conclusions}

\begin{sloppypar}
This paper presents ESPBench, a benchmark for data stream processing architectures in the enterprise context, where streaming data is combined with structured business data.
The ESPBench workload covers all core functionalities of DSPSs.
As part of ESPBench, we provide an example implementation using Apache Beam and a comprehensive toolkit, which simplifies the benchmark use.
The toolkit allows an objective result calculation, i.e., ESPBench does not rely on the different and differently measured performance metrics several systems provide.
We validated ESPBench using the example query implementation in an experimental evaluation, benchmarking three state-of-the-art DSPSs along with a modern DBMS.
The benchmark results that incorporate the impact of Apache Beam reveal that no system outperforms the others for all scenarios.
\end{sloppypar}

To include additional use cases for ESPBench in the future, the validator can be extended to allow ignoring the query result order, focusing only on the existence of results.
This additional feature allows for extended scaling concerning the Apache Kafka topic partitions.
In the current setting, we topics with a single partition to guarantees the correct order of records.
However, there might be use cases where the result order is not crucial that justify having this validator option.
By scaling via Apache Kafka topic partitions, the data input rate manageable by the message broker can be increased.

Future work further includes query extensions, e.g., other kinds of windowing scenarios, and extended evaluations.
Topics of interest include analyses of the scalability characteristics of the DSPSs and the impact of altered DBMSs or their  configurations on the latencies.
In this work, we have focused on open-source DSPSs.
Results of comparisons with commercial systems using ESPBench will lead to a more complete overview of the DSPS landscape.
Regarding the benchmark results, the metrics calculated by the validator and result calculator tool can be extended by, e.g., throughput.

ESPBench represents an easy-to-use yet meaningful benchmark that produces objectively determined results.
We are confident that ESPBench improves performance benchmarking in the area of DSPSs and invite others to apply it and to propose improvements.
\balance

\bibliographystyle{ACM-Reference-Format}
\bibliography{refs}


\end{document}